\documentclass[acmsmall,screen,nonacm]{acmart}

\usepackage{xcolor}
\usepackage{graphicx} 
\usepackage{listings}
\usepackage{subcaption}
\usepackage{lmodern}
\usepackage{cleveref}

\title{Byte-level Object Bounds Protection}
\author{Piyus Kedia}
\affiliation{%
  \institution{IIIT Delhi}
	\country{India}
}
\date{October 2025}

\begin{document}

\begin{abstract}
Low-level C programs remain highly vulnerable to out-of-bounds memory corruption. State-of-the-art precise defenses either introduce severe runtime overhead due to metadata memory lookups, or break standard C semantics by disallowing partial structs or the creation of an object's end address (EA), a legal operation ubiquitous in real-world C code. Conversely, practical alignment-based solutions achieve efficiency only by relaxing protected bounds.

We present PRISM, a precise, zero-lookup object-bounds scheme that eliminates these restrictions. PRISM compresses a 47-bit EA into the 17-bit unused tag area of a 64-bit pointer. By enforcing the invariant that a statically known starting address (KSA) cannot exceed the EA, PRISM completely eliminates the need for costly metadata memory fetches in nearly all bounds checks, while strictly retaining precise object bounds. Our invariant also simplifies the lower-bound checks in existing alignment-based solutions, thus improving their performance.

To achieve high throughput, PRISM introduces q-padding, an optimization that safely removes bounds checks for constant-offset accesses (such as struct fields) while maintaining precise, byte-level protection for the variable-indexed accesses primarily exploited by attackers.

Evaluated on SPEC 2017, PRISM achieves an arithmetic mean CPU overhead of 46.1\% with a 32-byte q-padding (dropping to 31.3\% in a 32-bit address space). On highly concurrent, real-world workloads, PRISM secures a fully saturated Apache web server with only an 11.1\% throughput reduction, demonstrating its readiness for production deployment. Furthermore, PRISM successfully detected an out-of-bounds violation in \texttt{gcc} that prior tools missed due to their lack of support for partial structs.
\end{abstract}

\maketitle

\newcommand{\secref}[1]{§\ref{#1}}

\section{Introduction}

Low-level C programs remain vulnerable to security attacks \cite{ropsecurity, jopsecurity, smashing-security, rop1security, breakingsecurity, missingsecurity, losingsecurity, guardsecurity} due to unchecked out-of-bounds accesses. Such bugs continue to appear in widely deployed software such as web servers \cite{nginx}, distributed memory caching \cite{memcached}, OpenSSL \cite{heartbleed}, GLIBC \cite{ghost}, the Windows Server Message Block protocol \cite{eternalblue}, showing that even critical, heavily reviewed code is not immune.

The existing techniques for spatial safety can be categorized into two main categories: object-bounds approaches \cite{jones,baggy,lfp1,lfp2,pari,sgxbounds,cguard,shadowbound,camp,cred,backward} that ensure accesses lie within the object's start and end, and the pointer-bounds approaches \cite{softbound,effective,austin,ccured,condit2003ccuredreal,cyclone,xu2004efficient} that maintain distinct bounds per pointer, often with higher memory overhead.
A third class \cite{address,valgrind,sanrazor,debloat,tail,purify,light,range} detects overflows and underflows but cannot prevent inter-object overflows. Hardware-based designs have also been proposed \cite{hardbound,mpx,mpxexplained,cheri,sparc_adi,arm_mte}.

Throughout the paper, we use {\em SA} and {\em EA} to denote the start and end addresses of an object, respectively, and {\em KSA} to denote a {\em statically known starting address}. These terms will appear frequently in our design and analysis.
We present PRISM, an object-bounds mechanism. A key challenge for such systems is efficiently computing SA and EA. One way of reducing the overhead of finding SA and EA is to relax the bounds of the objects.
Alignment-based schemes reduce this cost by rounding allocation sizes to the nearest $2^k$, and setting the alignment to the relaxed size. Due to the alignment property, they compute the SA from an internal address of the object by resetting the lower $k$-bits. The EA can be computed by adding $2^k$ to the SA. These remain the fastest software-only solutions.

Tag-based solutions embed metadata into the virtual address; we refer to the location of metadata as {\em tag area}. Precise-bounds systems \cite{sgxbounds,cguard,framer} store metadata in a tag area to find the SA or EA. \cite{cguard, framer} use 16-bit tag-area (corresponding to unused bits in X86\_64) to efficiently locate SA. The object size is stored just before the SA, needed to compute EA. SGXBounds \cite{sgxbounds} restricts the address space to 32-bit and directly embeds EA in the top 32-bit tag area. The SA is stored just after the EA. One common issue in all three approaches is that at least one memory access is needed to compute both bounds.

To break this memory wall, we present \textbf{PRISM}, a zero-lookup object-bounds architecture. PRISM introduces a novel heap layout and a 17-bit pointer tagging scheme that compresses a 47-bit EA directly into the unused upper bits of a 64-bit pointer. By ensuring that the KSA inherently lies within the inclusive range of \texttt{[SA, EA]}, PRISM can perform precise bounds checking without needing to compute or fetch the SA. Consequently, PRISM eliminates the metadata memory access in almost all cases. In the SPEC 2017 suite, \texttt{perlbench} required memory accesses for only 2.8\% of checks, while all other benchmarks required them less than 0.006\% of the time.

The second major overhead comes from the bounds check itself. CGuard \cite{cguard} introduced a size-invariant optimization: if the type of KSA is Ty and \texttt{KSA <= EA-sizeof(Ty)} then accesses in the range \texttt{[KSA,KSA+sizeof(Ty)]} need no dynamic checks.
This restriction is inherent to ShadowBound \cite{shadowbound}; \cite{cguard} permits violations via trap-based handling, but they modified SPEC benchmarks due to high trap costs. For each 8-byte aligned memory, \cite{shadowbound} uses 8-byte shadow memory to store the metadata, which is fetched to check the validity of every pointer arithmetic. \cite{shadowbound} cannot support storing and creation of EA (frequently used in iterators) because its 8-byte per-slot shadow memory cannot represent EA's bounds, which are needed when an in-bounds pointer is created from EA. Supporting EA with precise bounds checks requires a non-trivial redesign.

Similar to \cite{cguard}, \cite{shadowbound} ensures that all pointer escapes are within the inclusive range of \texttt{[SA, EA-sizeof(Ty)]}. For other pointer arithmetics, it checks that the resulting pointer is within the inclusive range of \texttt{[SA, EA-1]}. It does not protect the access size (loads and stores are not instrumented). \cite{shadowbound} adds additional padding to protect the access size, which at most can fall in the padded region. \cite{shadowbound} also skips bounds checks when two pointer expressions differ by a constant offset (e.g., {\tt arr[i]} and {\tt arr[i+10]}), by permitting accesses into a padded region. PRISM introduces {\em q-padding} optimization, which removes bounds checks for {\tt struct} accesses in many cases. Unlike other padding-based mechanisms \cite{shadowbound,baggy,lfp1,lfp2}, PRISM permits accesses to padded bytes only when the access is within a constant offset from KSA. This design choice ensures that all variable-indexed memory accesses remain precisely checked. This property is essential because, in real-world exploits \cite{nginx,memcached,heartbleed}, attackers manipulate variable indices to trigger overflows; PRISM therefore guarantees precise protection for these accesses. In the absence of precise checks, these bugs may remain undetected. If the q-padding is not used, PRISM provides precise protection for all heap, global, and stack accesses. By further relaxing the allocation bounds of the objects, the q-padding optimization also improves the performance of alignment-based solutions. PRISM enables applications to create and store EA, thereby eliminating the key limitation of \cite{shadowbound}. PRISM's bounds checking mechanism is much more efficient than \cite{cguard}.

Like \cite{framer, cguard}, PRISM categorizes objects into small and large. \cite{cguard} stores offset relative to SA in the tag-area, and updates them on pointer escapes, which adds additional overhead. FRAMER \cite{framer} avoids tag updates, but needs two memory accesses for large objects. PRISM uses frame-based encoding for small objects (as in \cite{framer}), but stores EA relative offset in the tag area (as in \cite{sgxbounds}). For large objects, the PRISM approach differs significantly from that of \cite{cguard, framer}. PRISM's design is motivated by the need to compute the EA effectively without needing to determine the category, small or large, of the object. By further restricting the address space to 32-bit, PRISM achieves significantly better performance than the alignment-based approaches, while protecting precise bounds.

To summarize, we make the following contributions.
\begin{itemize}
\item A novel heap layout that compresses 47-bit EA in the 17-bit tag area, enabling PRISM to compute EA without memory access.
\item A technique that removes memory accesses for nearly all bounds checks, and reduces lower-bound cost in alignment-based approaches by bypassing extra XOR operations.
\item An optimization that cleanly removes checks for constant-offset accesses while maintaining precise, byte-level protection for highly vulnerable variable-index accesses.
\item A 32-bit implementation that outperforms alignment-based approaches while enforcing precise object bounds.
\item A low-overhead scheme that supports all C programs that adhere to defined C semantics.
\item An LLVM-based implementation evaluated against SPEC 2017, Phoenix, and fully saturated Apache web servers---achieving as low as 11.1\% overhead on Apache---alongside security testing against real-world exploits.
\end{itemize}

\section{A faster design}
At a high level, PRISM computes the KSA by rolling back all pointer arithmetic, and it restricts pointer addresses to remain within the inclusive range [SA, EA]. Objects are allocated from fixed-alignment frames. For each object, PRISM stores (i) the offset of EA within the frame in the top 17-bit tag area, and (ii) the SA at location EA. During bounds checking, EA is reconstructed from KSA using the tag area; SA is fetched only when the checked pointer is less than KSA. Bounds checks are inserted before every memory access and whenever a pointer escapes its static scope (i.e., call, return, or store). In the following subsections, we elaborate on these details.

\subsection{KSA computation}
The KSA computation is similar to the static-base computation in CGuard \cite{cguard}. The KSA is computed by rolling back all the pointer arithmetic and typecasts. In an SSA form, a PHI instruction is generated to combine the values (possibly different) coming from different branches. A SELECT instruction is similar to a ternary if-else statement in C. For a PHI and SELECT instruction \texttt{I}, a new PHI and SELECT instruction, \texttt{NI}, is created that serves the KSA for \texttt{I}. The operands of \texttt{NI} are the KSAs of the operands of \texttt{I}. The new instructions are shown below. The pair \texttt{<ptr1, pred1>} indicates that the value (\texttt{ptr1}) is coming from the predecessor basic-block (\texttt{pred1}).

\begin{lstlisting}
I: PHI <ptr1, pred1>, <ptr2, pred2>, ..., <ptrn, predn>
NI: PHI <ksa(ptr1), pred1>, <ksa(ptr2), pred2>, ..., <ksa(ptrn), predn>

I: SELECT cond, ptr1, ptr2
NI: SELECT cond, ksa(ptr1), ksa(ptr2)
\end{lstlisting}

\subsection{KSA-invariant}
The KSA-invariant requires that a KSA always lies within the inclusive range [SA, EA]. PRISM adds dynamic checks to enforce this invariant at runtime. If the KSA-invariant is violated, the program triggers a runtime exception. This behaviour is fully compliant with the ANSI/ISO C standard, which treats the creation of pointers outside an object's bounds as undefined behavior. To enforce KSA-invariant, whenever a pointer leaves the static scope (i.e., passed to a function, stored in memory, or returned to a caller), we insert a bounds check to enforce the KSA-invariant property. Notice that after leaving the static scope, a pointer can become a KSA at other program points.

\subsection{Heap layout}

\begin{figure}
\centering
\begin{subfigure}[b]{0.4\columnwidth}
  \includegraphics[width=\linewidth, height=4.5cm]{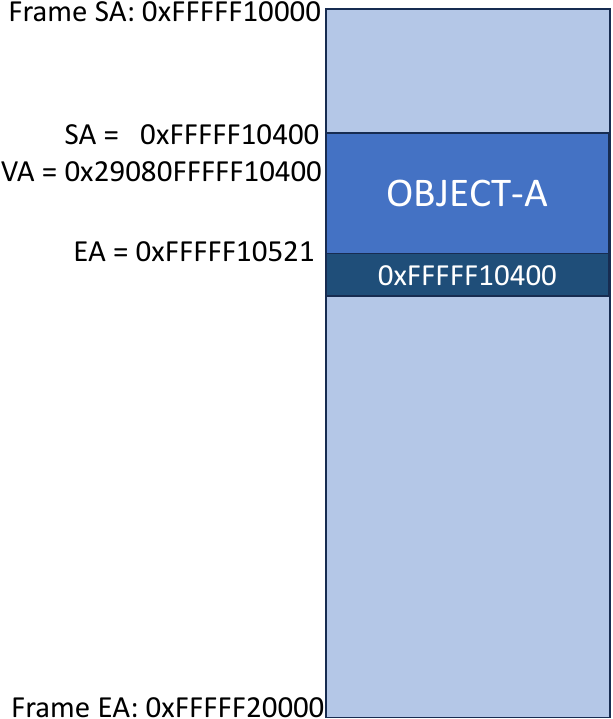}
  \caption{Layout of the small objects in the heap.}
  \label{fig:small}
\end{subfigure}%
\hfill%
\begin{subfigure}[b]{0.55\columnwidth}
  \includegraphics[width=\linewidth, height=4.5cm]{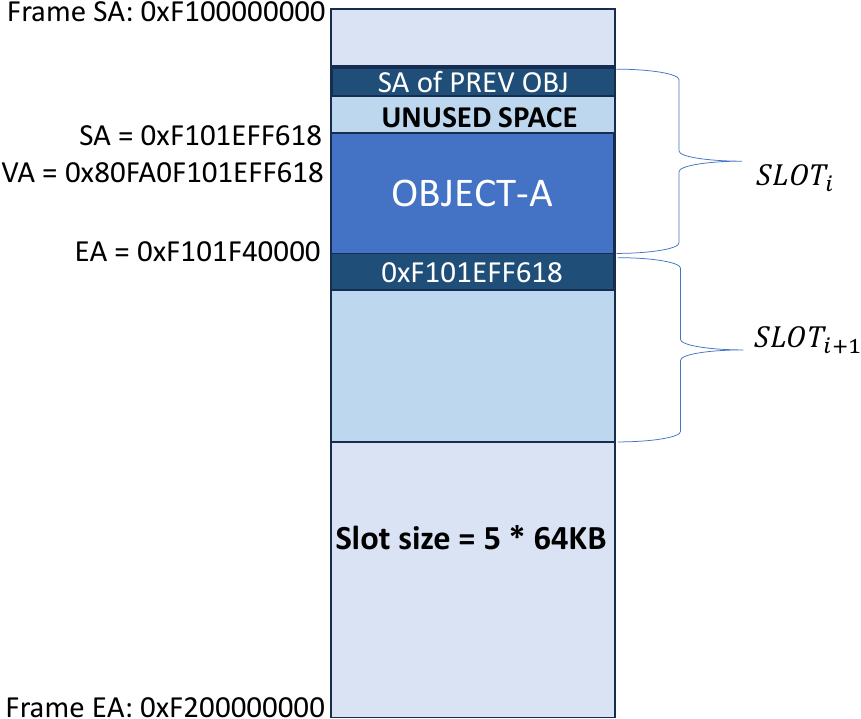}
  \caption{Layout of the large objects in the heap.}
  \label{fig:large}
\end{subfigure}
\caption{Heap layout}
\label{fig:layout}
\end{figure}

The heap maintains small (64 KB) and large (4 GB) frames, each aligned to its respective size. Each object reserves 8 bytes to store its SA at EA. Objects of size less than or equal to `64 KB - 8' bytes are called small objects and are allocated from a small frame. The other objects, referred to as large objects, are allocated from the large frames. The maximum size of an object is restricted to `4 GB - 64 KB'. A small frame is shown in Figure~\ref{fig:small}. Notice that the SA of Object-A is stored at its EA. The SA of the frame is aligned to 64 KB.

A large frame is divided into fixed-size slots of size k $\times$ 64 KB, where k is fixed per frame. A slot of size s can hold any object of size at most s - 8 bytes. The first 8 bytes of each slot store the SA of the object allocated in the {\em previous} slot. For an allocation request of x bytes from a slot beginning at address y, PRISM computes the object's SA as y + s - x, inserting s - x bytes of padding before the object. This padding guarantees that the object's EA is 64 KB-aligned, allowing PRISM to omit the lower 16 bits of EA from the metadata. The SA of the newly allocated object is stored in the first 8 bytes of the {\em next} slot, i.e., at the object's EA. Figure \ref{fig:large} illustrates this layout: the A's EA is aligned to 64 KB, its SA is stored at the start of the following slot, and additional padding is inserted because A's size is less than s - 8 to preserve the alignment property. The first eight bytes of A's slot store the EA of the object in the previous slot. The SA of the large frame itself is aligned to 4 GB.

For the alignment-based approach, we round the application request of \texttt{x} bytes up to the nearest $2^k$ value A, such that \texttt{A $\ge$ x + 1}. The alignment of the object is set to A. The bounds check in aligned mode ensures that at most \texttt{A - 1} bytes can be accessed after the SA; therefore, at least \texttt{x + 1} bytes are allocated to allow applications to create and store EA.

For 32-bit mode, we allocate eight additional bytes and store SA at memory location EA.

\subsection{Metadata and EA computation}
The top 17 bits of a user-mode virtual address are always zero in most operating systems running on top of x86\_64. Therefore, PRISM uses them as the tag area.
The top bit in the tag area stores the type of the frame, which is zero for a small frame and one for a large frame. The other bits are called the tag-offset.

\vspace{0.2em}
\noindent\textbf{Small frame:} For a small object, the metadata is EA's offset from the frame's SA. As the frame is 64 KB aligned, the lower 16 bits of the EA are the actual offset, which is stored in the tag-offset. Notice that, as KSA always belongs to [SA, EA], and thus always resides inside the frame, the EA can be computed by clearing the lower 16 bits to obtain the SA of the frame and adding the tag-offset, as shown at lines 5-7 in the compute\_ea routine below.

\vspace{0.2em}
\noindent\textbf{Large frame:} Large frames are 4 GB–aligned, so EA offsets are 32-bit values. However, EA is always 64 KB–aligned, which forces the lower 16 bits of the offset to be zero. PRISM, therefore, stores only the upper 16 bits of the EA offset in the tag-offset. At runtime, EA is computed from KSA by clearing the lower 32 bits to obtain the SA of the frame, left-shifting the tag-offset by 16 bits, and adding the shifted value to the frame's SA. The corresponding logic is shown at lines 10-12 below. At line 10, {\tt compute\_ea} directly computes the 32-bit offset by moving the tag area to bits 16-31 and applying a mask operation to clear other bits.

\begin{lstlisting}
// Input: known staring address, ksa
// Output: End address, ea
compute_ea(ksa) {
	if ((ksa & (1ULL << 63)) == 0) {
		tag_offset = ksa >> 47;
		frame_start = ksa & 0x7FFFFFFF0000;
		return frame_start + tag_offset;
	}
	else {
		frame_offset = (ksa >> 31) & 0xFFFF0000;
		frame_start = ksa & 0x7FFF00000000;
		return frame_start + frame_offset;
	}
}
\end{lstlisting}

For the alignment-based scheme, we encode the base-2 logarithm of the allocation size in the tag area, whereas for the 32-bit scheme, we store EA directly in the tag area.

\subsection{Bounds check}
Once EA is known, the bounds-check logic is identical for small and large objects. Before each memory access, PRISM clears the tag bits of the pointer and its KSA so that the check operates on raw addresses. The top 17 bits are removed from all arguments passed to {\tt bounds\_check} and {\tt bounds\_check32}.
Note that we need to do a memory access only if \texttt{ptr < ksa}. If we always perform memory access, the average overhead for SPEC benchmarks increases by 1.54\%, with a maximum of 4.38\% for mcf.

\begin{lstlisting}
// ksa: known starting address 
// ptr: address being accessed
// access_size: number of bytes being accessed
// ea: end address
bounds_check(ksa, ptr, access_size, ea) {
	// the eight byte staring address is stored at ea
	// the starting address is read only if ptr < ksa
	if (ptr + access_size > ea || (ptr < ksa && ptr < *(void**)ea)) {
		abort();
	}
}
\end{lstlisting}

The bounds check routine for the alignment-based approach is shown below. Following a similar approach to \cite{lfp1,lfp2}, we allocate one additional byte for the alignment-based approach to allow the application to create and store EA. The dynamic checks ensure that the pointer limit can't be greater than or equal to the EA obtained during the allocation, which is fine because the allocated EA is always greater than the actual EA. As stated earlier, most of the time the ptr $\ge$ ksa holds, checking ptr < ksa eliminates the need to execute an additional \texttt{XOR} operation in the lower bound. If we don't check ptr < ksa, we observed a 3.5\% increase in the average overhead of SPEC with a maximum of 10.15\% for \texttt{xz}. We never observed any slowdown due to this because the condition rarely fails.

\begin{lstlisting}
// tag: tag stored in the tag area of ksa
bounds_check_2k(ksa, ptr, access_size, tag) {
	aligned_size = (((size_t)1) << tag);
	if (  ((ksa ^ (ptr + access_size)) >= aligned_size) || 
	      (ptr < ksa && ((ksa ^ ptr) >= aligned_size))  ) {
		abort();
	}
}
\end{lstlisting}

\subsection{Pointers escape}
\label{sec:escape}
To ensure that the KSA-invariant holds, whenever a pointer escapes the static scope, i.e., stored in memory, passed to a function, or returned from a function, we add a bounds check with the access size set to zero. This ensures that the escaping pointer can at most be the EA. The bounds check is added if an escaping address and its KSA are not statically known aliases. In actual programs, sometimes a placeholder for an invalid address, such as \texttt{(void*)-1}, is used. When these pointers require dynamic checks, the bounds checks fail. Fortunately, in most cases, they are static aliases of their respective KSAs. In cases where static analysis can't precisely determine if two pointers are aliases, i.e., when used in PHI and SELECT, we perform a bounds check only if the escaping pointer is not equal to KSA at runtime. If pointer arithmetic is performed on an escaping placeholder address, the bounds checks fail; fortunately, we encounter very few such cases (discussed in \ref{sec:security}).

\subsection{Q-padding optimization}
In q-padding optimization, we allocate \texttt{q} additional bytes, and store SA at location \texttt{EA+q}. Because the existing checks ensure that KSA can at most be EA, we can safely eliminate the checks for the memory access that accesses at most \texttt{q}-bytes starting from KSA. The metadata and the EA computation logic remain unchanged. The bounds check logic is changed slightly as follows:

\begin{lstlisting}[numbers=none]
if (ptr + access_size > ea || (ptr < ksa && ptr < *(void**)(ea+q)) {
	abort();
}
\end{lstlisting}

For the alignment-based approach, the bounds check logic remains unchanged for q-padding. If the aligned allocation size is A without q-padding, we allocate \texttt{A+q-1} bytes with alignment A. The bounds check ensures that the KSA can be at most \texttt{SA+A-1}; therefore, we can safely access up to \texttt{q} bytes starting at KSA.

The q-padding significantly reduces the total number of bounds checks, but allows the application to safely access the padded area in the code path where checks are removed.
We found that up to 37\% of bounds checks are removed with only 8-byte padding (Section~\ref{sec:evaluation}). The average overhead for SPEC reduces by 9.51\% for the alignment-based approach and 32-byte padding. Interestingly, for q=32, the average overhead of the alignment-based approach is only 1.85\% lower than PRISM.

The linked list search and update routine illustrates why this optimization is highly effective.

\begin{lstlisting}
struct List {
	int key;
	int val;
	struct List *next;
};

search_and_update(struct List *head, int key, int newval) {
	while (head != NULL) {
		if (head->key == key) {
			head->val = newval;
			break;
		}
		head = head->next;
	}
}
\end{lstlisting}

In the above example, \texttt{search\_and\_update} iterates a linked list to find a node that stores \texttt{key}. If the search is successful, the \texttt{val} field is updated. Inside the loop, the variable \texttt{head} is KSA. Actually, \texttt{head} is a PHI instruction in the SSA form, and the corresponding KSA is also a PHI instruction. Both PHI instructions are aliases, so \texttt{head} is treated as KSA.

At line 9, 4 bytes are accessed starting at \texttt{head}, so for q=4, this bounds check is not needed. At line 10, 4 bytes beginning at offset four are accessed, so for q=8, this check is removed. At line 13, 8 bytes starting at offset eight are accessed; therefore, the check can be removed if q $\ge$ 16.
For q=16, this routine will not have any bounds checks. This example demonstrates that q-padding optimization is highly effective, particularly for {\tt structs}.
\cite{cguard,shadowbound} size-invariant optimization can also eliminate all the checks in this example; however, it requires {\tt head} to point to a memory location that is at least \texttt{sizeof(struct List)} long, which limits its applicability, because of the need to change the source code if this property doesn't hold. In fact, an out-of-bounds violation bug exists in a partially allocated {\tt struct} in \texttt{gcc} benchmark (discussed in Section~\ref{sec:security}).

\subsection{Stack and global accesses}
The sizes of the stack and global variables are statically known, and therefore, we don't need to add metadata for bounds checking. However, if a stack or global address escapes the local scope, i.e., is stored in a memory location, passed to a function, or returned from a function, we can't precisely distinguish the escaped address from a heap address.
For the escaped stack and global addresses, we allocate additional memory and padding (if q > 0), store SA at location EA + q, similar to the heap objects. Finally, we replace the escaping pointers with another pointer that also stores the correct metadata. CGuard \cite{cguard} handles stack and global escapes in a similar way.

\subsection{Pointer comparison and subtraction}
As we add metadata to the escaping stack and global addresses, two versions of the same address are possible: one with and one without the metadata. To correctly handle this, we instrument all pointer comparison and subtraction operations in which one operand is a global variable or an escaped stack variable. In these operations, we reset the metadata from the pointer operands before the operation. Note that \cite{cguard} instruments all pointer comparisons and subtractions, because the metadata are frequently updated for all types of objects.

\subsection{Library call}
We don't instrument library calls. This creates a problem because the library functions can dereference the tagged pointers. Moreover, some library functions also return raw pointers, which causes our dynamic checks to fail. We use the \cite{cguard} approach, which involves using wrapper functions to clear and embed metadata when the library routines are called.

\section{Implementation}
We extended the publicly released artifact \cite{cguardimp} of CGuard \cite{cguard} to implement PRISM. \cite{cguardimp} provides support for wrapper functions for a wide range of library calls utilized by the benchmarks. It also had support for several optimizations that we could directly reuse. \cite{cguardimp} extended the LLVM \cite{llvm} compiler and JEMALLOC \cite{jemalloc} allocator. For 32-bit mode, we used a custom GLIBC-2.38 library.

\subsection{32-bit support}
To support 32-bit addresses, we modified GLIBC's \texttt{mmap} routine. Along with the user-supplied flag, we pass the \texttt{MAP\_32BIT} flag to ensure that the returned address lies within the top 2GB of the address space. If this allocation fails, we retry within the 2 to 4 GB address space using \texttt{MAP\_FIXED\_NOREPLACE}. These implementations are sufficient to run all the benchmarks. We also introduce a custom 32-bit stack. At the beginning of \texttt{main}, we call an initialization routine that allocates a new stack using \texttt{mmap} and switches execution to it.

\subsection{Stack allocations}
If a stack address escapes the static scope, we attach metadata to the escaping address in the same way as for heap objects. If the object size is less than 64 KB, we must ensure that the object does not straddle a small-frame boundary at runtime. To guarantee that the object remains entirely within the small frame, we set its alignment to the nearest power of two, \texttt{z}, such that \texttt{z $\ge$ object\_size + 1}. For large objects, we set the alignment to 64 KB, promote the size to a multiple of 64 KB, and insert padding before SA so that EA is aligned to a 64 KB boundary. In both cases, we store SA in the additional eight bytes following EA.

For dynamic alloca, it's challenging because we only know the size of the object at runtime. Therefore, for these objects at runtime, if the object doesn't stay entirely within the small frame, we promote the size to a multiple of 64 KB, set the alignment of the object to 64 KB, and insert padding before the SA so that EA is aligned to 64 bytes. The corresponding pseudocode is shown below. As before, SA is stored in the eight bytes immediately after EA.

\begin{lstlisting}
ptr = alloca(size+8)  // original ptr = alloca(size)
if ((ptr ^ (ptr+size)) >= 64KB) {
	asize = Align(size, 64KB);
	ptr = alloca_aligned(asize+8, 64KB);
	ptr += (asize-size);
}
*((void**)(ptr+size)) = ptr;
\end{lstlisting}

For the alignment-based scheme, we add instrumentation to compute the nearest $2^k$ size, create a new {\tt alloca} with twice the size of $2^k$ as an argument, use pointer arithmetic to align the result to $2^k$, and replace the aligned result with the original usage.

\subsection{Global allocations}
For global variables, the size is known at compile time. If the size is greater than or equal to 64 KB, we set the alignment to 64 KB and add padding before the SA. SA is stored after at location EA. We ensure that small objects remain within a small frame by controlling the alignment of the data section and inserting padding on the boundary of the 64 KB frames in the data section, if needed.

If a global variable field is initialized with the address of a global variable, we store the tagged address during initialization, as in \cite{cguard}. One problem here is that we need to perform an `and' operation on the EA to compute the metadata, but currently, LLVM doesn't support the `and' operation in the initializer. Therefore, along with the address say {\tt x} in the initialization, we also store the offset in the tag area that can be added to {\tt x} to compute the EA. We changed the LLVM linker (lld) to compute and store the correct metadata at link time. To store the offset during compilation, we use a large tag area to support large objects; later, our custom linker stores the correct tag during linking.

Another issue is the external global variables whose size is unknown during compile time. To compute the metadata for external variables, for each exported global variable \texttt{gv}, we create a global alias \texttt{gv.ea} that points to the EA of the object. The size of an exported global variable \texttt{gv.ea - gv} is used along with \texttt{gv.ea} by our modified linker to compute the tag value.

Additionally, whenever a global variable escapes the static scope, we compute and add the metadata to the escaping address. We use \texttt{gv.ea} and \texttt{gv} to compute the metadata for the escaping imported global variables.

\subsection{Metadata reset}
Our bounds check logic doesn't use the metadata in KSA, EA, and the pointer address, say ptr. This simplifies the loading of SA from EA for the lower bounds. For the correct bounds check, we need to ensure that all pointer arithmetic operations performed to compute ptr from KSA are performed on values that don't contain the metadata.

Therefore, we use the following strategy for resetting the metadata of a pointer, denoted using instruction I.

\begin{lstlisting}
reset(I)
	if (I is a pointer arithmetic or bitcast)
		NI = I.clone()
		NI.setPointerOperand(reset(I.getPointerOperand()))
		return NI
	else if (I is phi or select)
		NI = I.clone()
		foreach pointer operand op in I
			NI.replaceOperand(op, reset(op))
		return NI
	else
		return CreateAnd(I, 0x7FFFFFFFFFFF);
\end{lstlisting}

We create a new instruction for every pointer arithmetic, bitcast, phi, or select instruction that was used to compute I from the KSA.
If I is a pointer arithmetic or a bitcast operation, we replace the pointer operand in the new instruction with the masked version of the pointer operand, which is recursively computed using the same algorithm (the actual implementation uses a map to reuse the previously computed masked version of the pointer, if it exists).
For phi and select, we create another phi and select whose operands are masked. If we reach KSA, we clear the top 17 bits to obtain the masked version.

Another advantage of this strategy is that it merges multiple reset operations on pointers derived from the same KSA.

\subsection{Optimizations}
We use several optimizations from CGuard \cite{cguardimp,cguard,cguardar}.
\subsubsection{Loop optimization:}
For array accesses in a loop, if \cite{cguard} statically knows the lower and upper bounds of the induction variable, and the array access postdominates the loop header -- it moves the bounds check outside the loop. Notice that the overall access size within the loop may not be a constant, and therefore, in the bounds check, it aborts the program if the pointer limit is less than the pointer to prevent a potential overflow. We used the same optimization and modified our bounds check logic. This optimization improves PRISM’ performance by 7.55\%, with x264 (34.6\%), imagick (10.5\%), and perlbench (7.42\%) seeing the largest gains.

\subsubsection{Combining bounds check:}
\cite{cguardar} combines two bounds checks if they are guaranteed to execute together and differ by a common offset. The optimization is shown below.

\begin{lstlisting}
Without optimization:
bounds check for [&arr[i+10], &arr[i+11]]
arr[i+10] = x;
bounds check for [&arr[i-10], &arr[i-9]]
arr[i-10] = y;
With optimization:
bounds check for [&arr[i-10], &arr[i+11]]
arr[i+10] = x;
arr[i-10] = y;
\end{lstlisting}

Even in cases where these accesses are not guaranteed to execute together, the actual implementation always removes the second check if the first check dominates the second check and the access range of the second check is completely covered by the first check.

Notice that the typical access size in a bounds check is eight bytes; however, due to this optimization, the access size could be arbitrarily large, e.g., consider the case when arr[i+1] and arr[i+10000000000] are combined. In such a case, there is a risk that the upper bound of the memory access can overflow the 64-bit virtual address space.
Therefore, similar to \cite{cguardar}, we set the 4MB threshold on the maximum constant value to perform this optimization, and to prevent overflow, we disallow access to the first 4MB of the process address space.

Because of the KSA-invariant, we could further optimize two checks of the form [ksa+c1, ksa+c2] and [ksa+c3, ksa+c4]. Here, c1>=0, c3>=0, c2>=c1, c4>=c3.
Note that because KSA is in the range [SA, EA], PRISM can ignore c1 and c3 from the bounds check. The bounds checks are performed on [ksa, ksa+c2] and [ksa, ksa+c4]. If the first check dominates the second check and c2>=c4, then the second check is removed. Otherwise, if both checks are guaranteed to execute together (i.e., the first check dominates the second check and the second check postdominates the first check), the first check is replaced by [ksa, ksa+c4] and the second check is removed.

\subsubsection{Elimination of lower bound check:} If the bounds check is performed on [ksa+c1, ksa+c2], where c1>=0 and c1<=c2<=1GB, then we remove the lower bound check because the starting address of the memory access is guaranteed to be greater than or equal to KSA. Note that the KSA always stays within the 47-bit address space. We can also pick a larger value, but 1GB was sufficient to cover all cases in the benchmarks. We used the existing implementation in \cite{cguardimp}. This optimization reduced the overhead for SPEC by 1.72\%.

\section{Evaluation}
\label{sec:evaluation}
We conducted our experiments on an eight-core Intel i9-9900K CPU with a 3.6 GHz processor, with 32 GB of RAM, a 512 GB SSD, and an RTL8111/8168/8411 PCIe Gigabit Ethernet Controller, running Linux 6.10.5-061005-generic. All applications are compiled at O3. We used the ``x86-branches-within-32B-boundaries'' compiler flag to mitigate the hardware bug (present on our CPU) reported in \cite{ierr}. Hyperthreading, swapping, and address randomization are disabled, and the CPU frequency governor is set to {\em performance}. Networking is disabled while running SPEC and Phoenix. This configuration ensures consistent results across multiple runs at different times. For Apache experiments, we directly connect our system to another machine with 1 Gbps NIC.

Throughout this section, PRISM, Pow2, and PRISM32 refer to the unpadded versions of our base approach, the alignment-based approach, and the 32-bit approach, respectively. PRISM(q), Pow2(q), and PRISM32(q) denote their q-padded variants. We report the arithmetic mean overhead. For memory overheads, we use the ``Maximum resident set size'' reported by /usr/bin/time.

We ran SPEC CPU 2017 \cite{spec20171, spec20172} with the reference input set. For Phoenix-2.0 \cite{phoenix}, we use the same configuration as \cite{cguard}, i.e., 2000x2000 for matrix-multiply, 3000x3000 for PCA, and 200000 for kmeans benchmarks, since the default large inputs run too quickly. Following \cite{cguard}, we use the pthread versions of kmeans, pca, and histogram, which run faster than their map-reduce version for the same input size. For scalability experiments, we disable additional CPU cores using the Linux CPU hotplug.

We use the median of five runs to compute the runtime of each benchmark. We use Apache-2.4.46, Apr-1.7.0, and Apr-util-1.6.1 in our experiments. For security evaluation, we use the BugBench \cite{bugbench} benchmark suite.

\subsection{SPEC CPU 2017 CPU overhead}
\begin{figure}[t]
  \centering
  \begin{subfigure}[t]{0.5\textwidth}
    \centering
    \includegraphics[width=\linewidth,height=3.5cm]{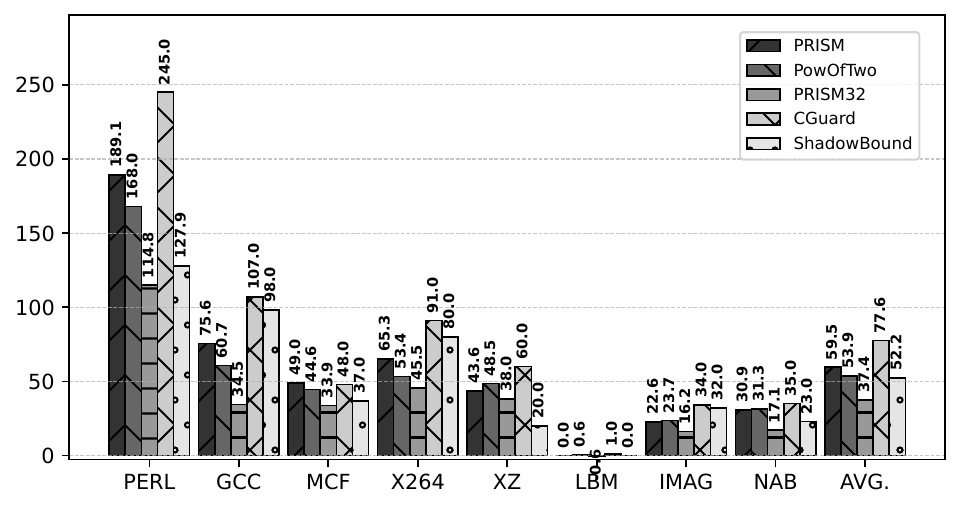}
    \caption{\% increase in execution time for PRISM, Pow2, PRISM32, CGuard, and ShadowBound.}
    \label{fig:spec_cpu1}
  \end{subfigure}
  \hfill
  \begin{subfigure}[t]{0.48\textwidth}
    \centering
    \includegraphics[width=\linewidth,height=3.5cm]{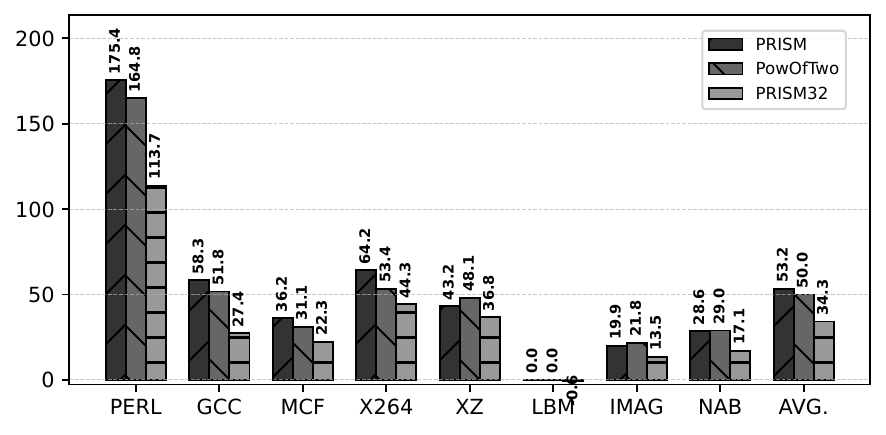}
    \caption{\% increase in execution time for PRISM(8), Pow2(8), and PRISM32(8).}
    \label{fig:spec_cpu2}
  \end{subfigure}
	 \vspace{1em} 
  \begin{subfigure}[t]{0.48\textwidth}
    \centering
    \includegraphics[width=\linewidth,height=3.5cm]{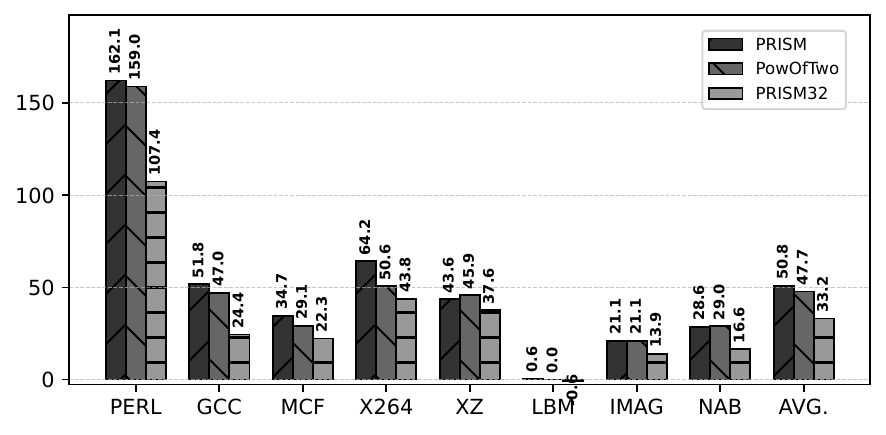}
    \caption{\% increase in execution time for PRISM(16), Pow2(16), and PRISM32(16).}
    \label{fig:spec_cpu3}
  \end{subfigure}
  \hfill
  \begin{subfigure}[t]{0.48\textwidth}
    \centering
    \includegraphics[width=\linewidth,height=3.5cm]{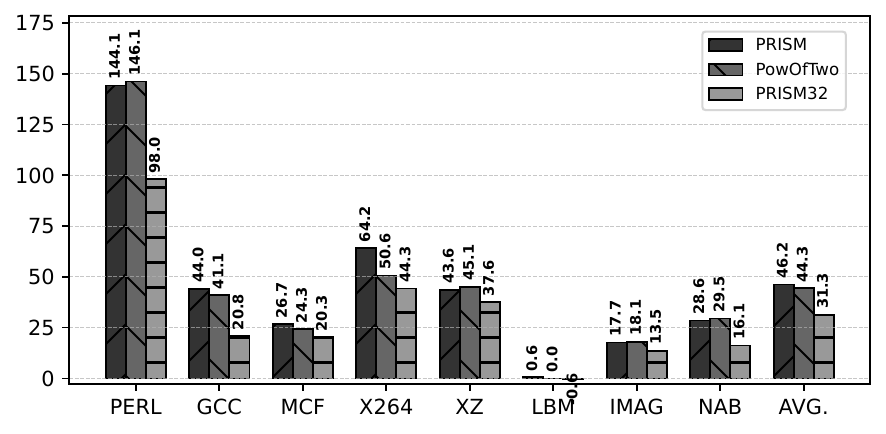}
    \caption{\% increase in execution time for PRISM(32), Pow2(32), and PRISM32(32).}
    \label{fig:spec_cpu4}
  \end{subfigure}
  \caption{CPU overhead of SPEC benchmarks.}
  \label{fig:spec_cpu}
\end{figure}

\begin{figure}[t]
  \centering
  \begin{subfigure}[t]{0.35\textwidth}
    \centering
    \includegraphics[width=\linewidth,height=3.5cm]{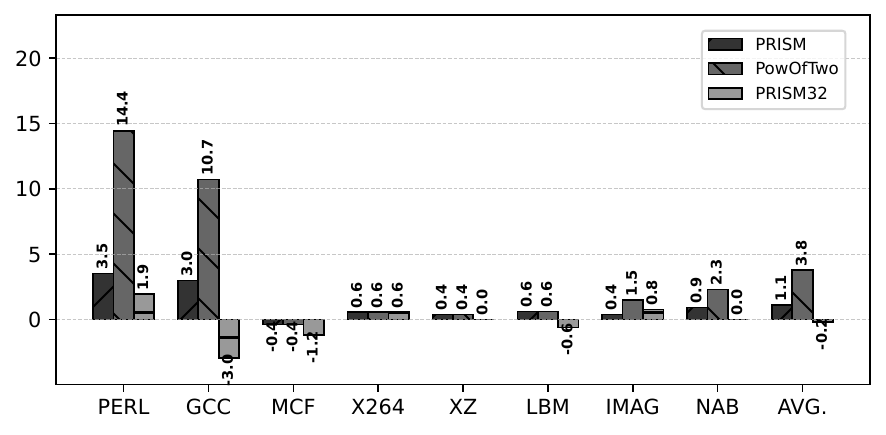}
    \caption{\% increase in execution time w.r.t. native due to the allocator changes for PRISM, Pow2, and PRISM32}
    \label{fig:allocator}
  \end{subfigure}
  \hfill
  \begin{subfigure}[t]{0.28\textwidth}
    \centering
    \includegraphics[width=\linewidth,height=3.5cm]{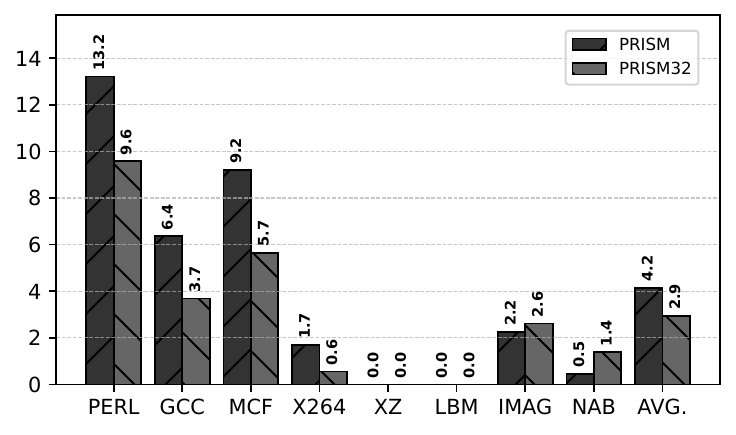}
		\captionsetup{width=0.8\textwidth}
    \caption{\% overhead of resetting the metadata for PRISM and PRISM32}
    \label{fig:ptrmask}
  \end{subfigure}
  \hfill
  \begin{subfigure}[t]{0.35\textwidth}
    \centering
    \includegraphics[width=\linewidth,height=3.5cm]{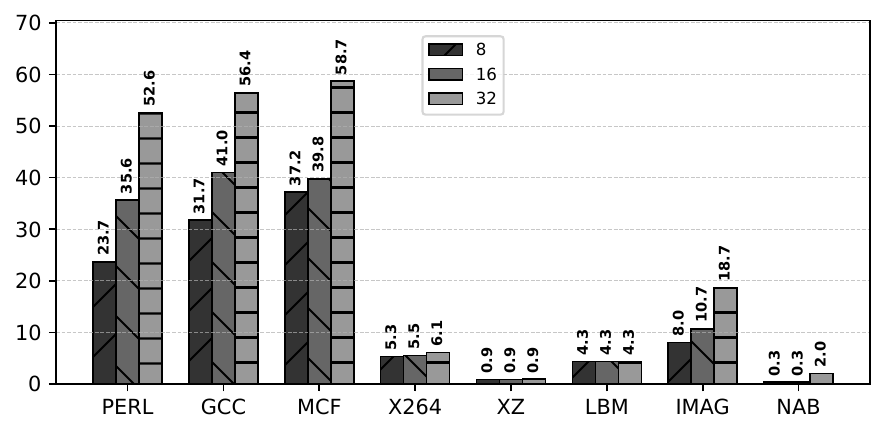}
    \caption{\% decrease in number of bounds check at runtime w.r.t. for q=8, 16, and 32}
    \label{fig:num_checks}
  \end{subfigure}
  \caption{Other interesting statistics.}
\end{figure}

To measure the CPU overhead of PRISM, we run SPEC 2017 benchmarks. Figure~\ref{fig:spec_cpu1} shows the percentage increase in execution time for PRISM, Pow32, PRISM32, CGuard \cite{cguard}, and ShadowBound \cite{shadowbound}. To ensure a fair comparison with our base approach, we use ShadowBound numbers with Runtime Driven Checking Elimination (rt-elim) disabled, as rt-elim relaxes the object bounds during checks, thereby not providing precise protection. We used an approximate value from the bar graph from their paper, because ShadowBound doesn't report absolute values. For CGuard, we used the default configuration that includes size-invariant optimization.

The average overhead of PRISM is 59.5\% compared to the 52.2\% overhead of ShadowBound. ShadowBound performs better for {\tt nab}, {\tt xz}, {\tt mcf}, and {\tt perlbench} (PERL in the figure). For the remaining benchmarks, PRISM performs better. Our 32-bit implementation could outperform ShadowBound for all except {\tt xz}. One reason behind this is that rt-elim doesn't entirely disable size-invariant optimization, as it doesn't instrument load and store operations. If a load or store is directly performed without pointer arithmetic, or a pointer arithmetic is used to compute the address of the first field (offset zero) of a {\tt struct}, no dynamic checks are performed. The second reason is that, unlike ShadowBound, all three versions of our implementation fully protect the stack and global accesses, and these checks can't be disabled because once an address escapes the local scope, PRISM can classify it as a heap or non-heap object.

PRISM' average overhead (59.51) is higher than Pow2's (53.85). This is mainly due to the higher cost of EA reconstruction, as PRISM must distinguish between the large and default objects. However, PRISM performs a simple upper bound check (a single `cmp'), whereas Pow2 requires an `xor' followed by `cmp'. For the lower bound, PRISM (memory access) and Pow2 (`xor') perform the extra work only when the pointer is less than KSA, which is extremely rare ($<=$ 2.81\% for perlbench; $<=$ 0.006\% for others). Because of a simpler check sequence, PRISM outperforms Pow2 for xz.

EA computation in PRISM32 is faster because all the objects belong to the same class. Consequently, PRISM32 outperforms Pow2 for all benchmarks, with an average overhead of 37.41\%, which is substantially lower than both Pow2 and PRISM.

Except for mcf, PRISM outperforms the default configuration of CGuard. CGuard needs to update the metadata on pointer escapes, and it perform a memory access to compute the EA; PRISM avoids both these cases.

To understand why PRISM' overhead is close to CGuard on {\tt mcf\_r}, we closely examine the {\tt cost\_compare} routine, which accounts for 23.7\% of the native runtime computed using the perf tool.

\begin{lstlisting}
int cost_compare(struct basket **b1, struct basket **b2 )
{
 if( (*b1)->abs_cost < (*b2)->abs_cost )
   return 1;
 if( (*b1)->abs_cost > (*b2)->abs_cost )
   return -1;
 if( (*b1)->a->id > (*b2)->a->id )
   return 1;
 else
   return -1;
}
\end{lstlisting}

The above routine requires four bounds checks at line 3 -- two for dereferencing \texttt{b1} and \texttt{b2} and two for reading \texttt{abs\_cost}. At line 7, field {\tt a} of \texttt{(*b1)} is accessed, but because the offset of {\tt a} in \texttt{struct basket} is zero, which is lower than that of {\tt abs\_cost}, and the bounds check for {\tt abs\_cost} already dominates this program point, another check with a lower offset is not required. The dereferences of \texttt{(*b1)->a} and \texttt{(*b2)->a} for {\tt id} require two more checks at line 7. Therefore, six bounds checks are required in this routine.

The CGuard default approach ensures that {\tt b1} and {\tt b2} point to a memory location that is at least eight bytes long. Therefore, two checks for dereferencing {\tt b1} and {\tt b2} at line 3 are eliminated. It also ensures that the loaded value of {\tt b1} and {\tt b2} at line 3 is at least {\tt sizeof(struct basket)} long, which is 32 bytes, and therefore, checks for dereferencing {\tt abs\_cost} are not needed. Similarly, when {\tt a} is loaded at line 7 {\tt (*b1)->a}, CGurad ensures that the loaded value is pointing to at least a 72-byte long memory area, because the type of the loaded pointer is {\tt struct arc*}, which is 72-byte long. Because {\tt id} is a member of {\tt struct arc}, a dynamic check is not needed. Ultimately, no check is needed for this routine in the CGuard approach.

With PRISM(8), the dynamic checks for dereferencing b1, b2 at line 3 and field {\tt id} at line 7 are removed, because the access range is within eight bytes of their respective KSAs. With q=24, the remaining checks for accessing {\tt abs\_cost} are removed because the offset of {\tt abs\_cost} is 16 and it's eight bytes long.
The size-invariant optimization can remove checks for {\tt struct}, which are hundreds of bytes long without additional padding. PRISM' goal is to support precise bounds while allowing all programs that adhere to the C standard, which allows EA and partial allocation of {\tt struct}.

Figures~\ref{fig:spec_cpu2}, \ref{fig:spec_cpu3}, and~\ref{fig:spec_cpu4} show the CPU overhead with 8, 16, and 32-byte padding. As expected, the overhead of {\tt mcf\_r} decreases from 49\% to 26.7\% with 32-byte padding for PRISM. The overhead with Pow2 also decreases from 44.6\% to 24.3\%. Interestingly, with 32-byte padding, the average overhead of PRISM is 46.2\%, compared to 44.3\% overhead for Pow2.

Figure~\ref{fig:num_checks} shows the percentage decrease in the total number of bounds checks at runtime with different padding sizes. With 32-byte padding {\tt perlbench}, {\tt gcc}, and {\tt mcf} show more than 50\% reduction in the dynamic checks. For perlbench, we overhead decreases by 45\%, 22\%, and 17\% with PRISM(32), Pow2(32), and PRISM32(32), respectively. For gcc, the reductions are 32\%, 20\%, and 14\%. For mcf, the reductions in overheads are 22\%, 20\%, and 14\%. These results indicate that the padding effectively reduces CPU overhead.

Figure~\ref{fig:allocator} shows the increase in the execution time due to the changes in the allocator. PRISM incurs 3.5\% and 3\% overhead for perlbench and gcc, caused by additional 8-byte padding for storing the metadata. Under Pow2 approach, the overheads rises to are 14.4\% and 10.7\%, driven by the stricter alignment requirement constraints. We profiled perlbench using the perf tool and found that it spent up to 17\% of the time in jemalloc, compared to 4\% during the native run, across various configurations used during the reference run. Interestingly, with 32-byte padding, the overhead of perlbench and gcc dropped slightly to 13.67\% and 9.52\%, potentially due to the change in allocator caching behavior.
We observed a 3\% reduction in the CPU overhead of gcc with PRISM32. This is because we use the 32-bit paging scheme in ggc-alloc.c, which turns out to be slightly faster than the 64-bit version (see Section~\ref{sec:security}).

Figure~\ref{fig:ptrmask} shows the additional CPU overhead of resetting the top 17 bits during memory accesses, excluding the allocator's overhead for PRISM and PRISM32. PRISM32 incurs lower overhead because the compiler can often use a 32-bit subregister to obtain the untagged pointer.

\subsection{SPEC memory overhead}
\begin{figure}[t]
  \centering
  \begin{subfigure}[t]{0.5\textwidth}
    \centering
    \includegraphics[width=\linewidth,height=3.5cm]{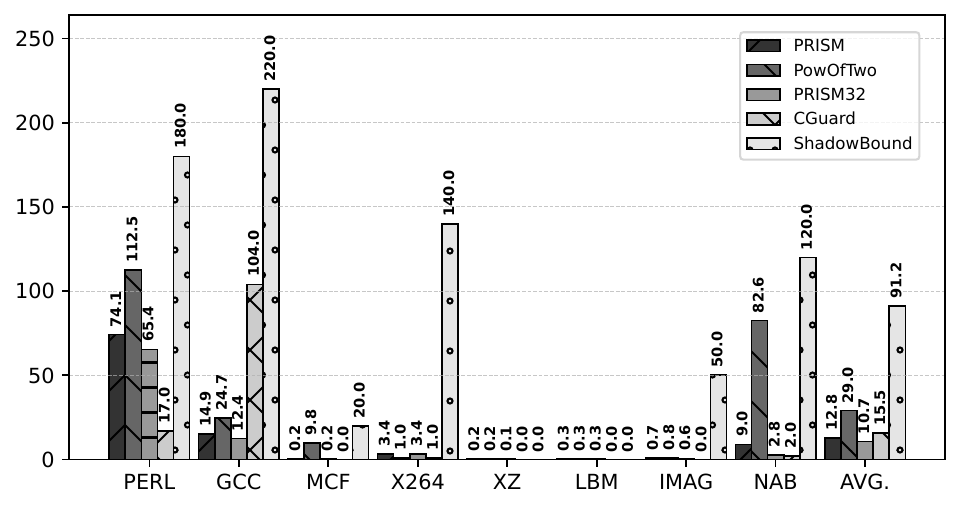}
    \caption{\% increase in peak memory consumption for PRISM, Pow2, PRISM32, CGuard, and ShadowBound.}
    \label{fig:spec_mem1}
  \end{subfigure}
  \hfill
  \begin{subfigure}[t]{0.48\textwidth}
    \centering
    \includegraphics[width=\linewidth,height=3.5cm]{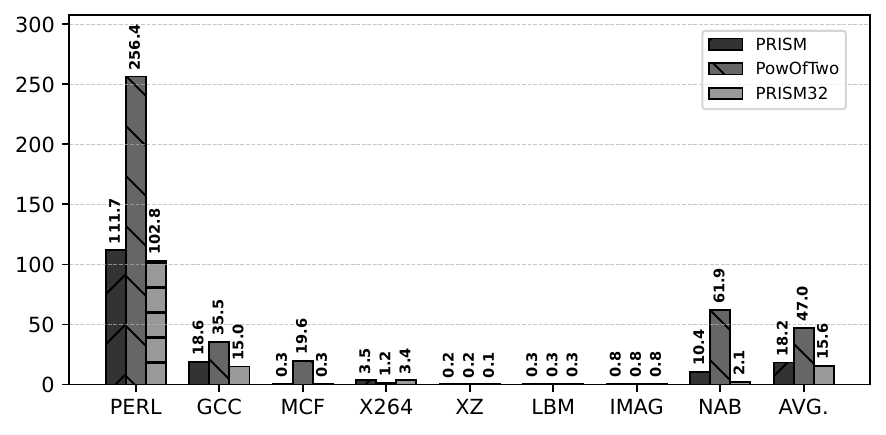}
    \caption{\% increase in peak memory consumption for PRISM(32), Pow2(32), and PRISM32(32).}
    \label{fig:spec_mem2}
  \end{subfigure}
  \caption{Memory overhead of SPEC benchmarks.}
  \label{fig:spec_mem}
\end{figure}

Figure~\ref{fig:spec_mem1} and Figure~\ref{fig:spec_mem2} show the memory overhead of SPEC without padding and with 32-byte padding. The maximum memory overhead is for the perlbench, in both cases. This is because we allocate an additional 40 bytes at an allocation site to prevent the benchmark from creating an out-of-bounds address (discussed in Section~\ref{sec:security}). When we ran the native version on the modified version of perlbench, we observed 54.01\% memory overhead compared to the unmodified version. This suggests that high overhead is primarily due to changes in the source code. As expected, the memory overhead is higher for the Pow2 approach. The memory overhead for perlbench jumped to 256\% from 112\% for Pow2(32). For PRISM32(32), the memory overhead for perlbench roughly doubled with respect to the unpadded version. Interestingly, for nab, the overhead of Pow2(32) is 62\% compared to 82\% for the unpadded version. We found that jemalloc treats objects of size 16384 or higher as large objects, which are immediately reclaimed after being freed. The reclamation of small objects is slightly delayed as they may end up in the per-thread caches. Out of all small objects, 57\% of the objects were of size 8192. With 32-byte padding and 8192-byte alignment, jemalloc allocates them from the 16384-byte bucket, classifying these as large objects. Consequently, these objects got reclaimed quickly, resulting in smaller peak memory overhead. CGuard reports 17\% and 107\% overheads for perlbench and gcc. The high overhead of gcc is mainly because of the changes related to size invariant. ShadowBound incurs high memory overhead due to its 2x memory requirement.

\subsection{Multithreading}

\begin{figure}[t]
  \centering
  \begin{subfigure}[t]{0.5\textwidth}
    \centering
    \includegraphics[width=\linewidth,height=2.5cm]{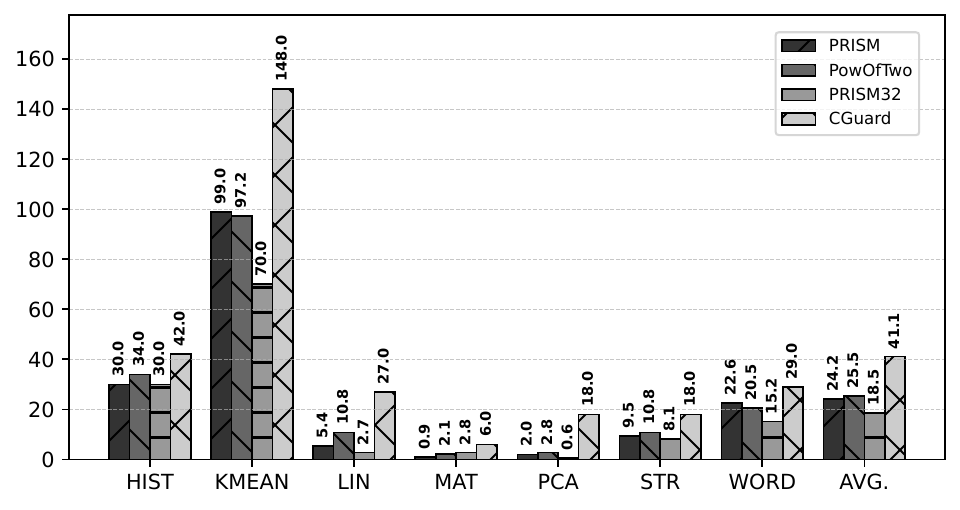}
    \caption{\% increase in execution time on a single processor for PRISM, Pow2, PRISM32, and address sanitizer without padding.}
    \label{fig:phoenix_cpu1}
  \end{subfigure}
  \hfill
  \begin{subfigure}[t]{0.48\textwidth}
    \centering
    \includegraphics[width=\linewidth,height=2.5cm]{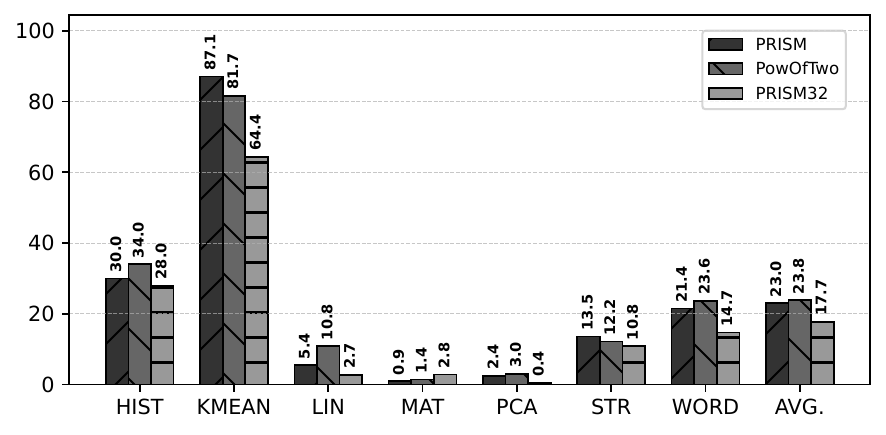}
    \caption{\% increase in execution time on a single processor for PRISM, Pow2, and PRISM32 with q=8.}
    \label{fig:phoenix_cpu2}
  \end{subfigure}


	 \vspace{0.3em} 
  \begin{subfigure}[t]{0.48\textwidth}
    \centering
    \includegraphics[width=\linewidth,height=2.5cm]{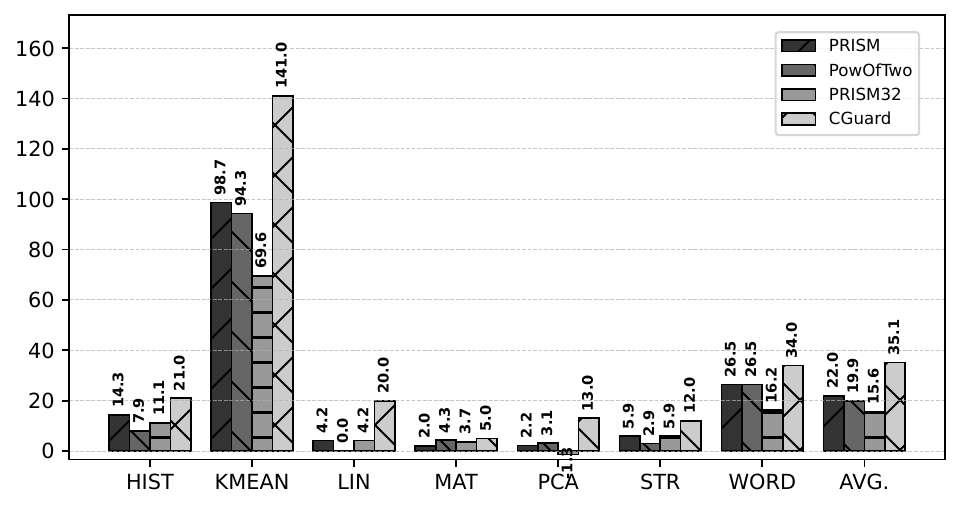}
    \caption{\% increase in execution time on four processors for PRISM, Pow2, PRISM32, and CGuard.}
    \label{fig:phoenix_cpu5}
  \end{subfigure}
  \hfill
  \begin{subfigure}[t]{0.48\textwidth}
    \centering
    \includegraphics[width=\linewidth,height=2.5cm]{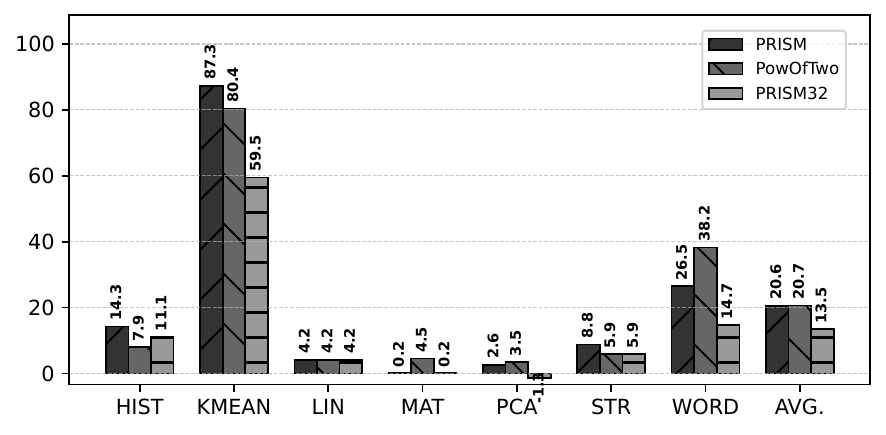}
    \caption{\% increase in execution time on four processors for PRISM(8), Pow2(8), and PRISM32(8).}
    \label{fig:phoenix_cpu6}
  \end{subfigure}
	 \vspace{0.3em} 
  \begin{subfigure}[t]{0.48\textwidth}
    \centering
    \includegraphics[width=\linewidth,height=2.5cm]{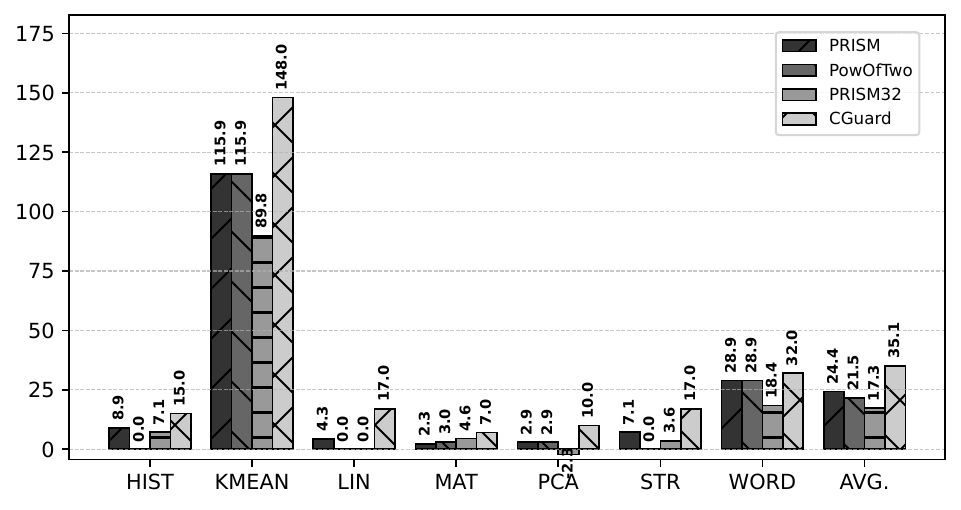}
    \caption{\% increase in execution time on eight processors for PRISM, Pow2, PRISM32, and CGuard.}
    \label{fig:phoenix_cpu7}
  \end{subfigure}
  \hfill
  \begin{subfigure}[t]{0.48\textwidth}
    \centering
    \includegraphics[width=\linewidth,height=2.5cm]{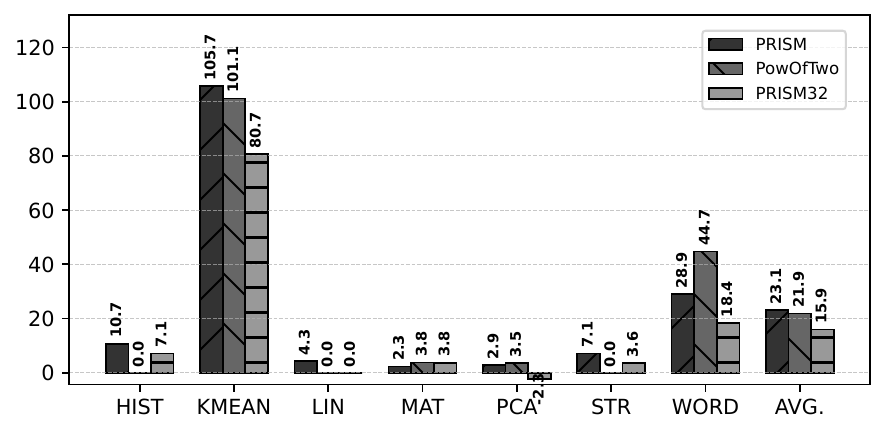}
    \caption{\% increase in execution time on eight processors for PRISM(8), Pow2(8), and PRISM32(8).}
    \label{fig:phoenix_cpu8}
  \end{subfigure}
  \caption{CPU overhead for Phoenix.}
  \label{fig:phoenix_cpu}
\end{figure}

\begin{figure}[t]
  \centering
  \begin{subfigure}[t]{0.5\textwidth}
    \centering
    \includegraphics[width=\linewidth,height=2.5cm]{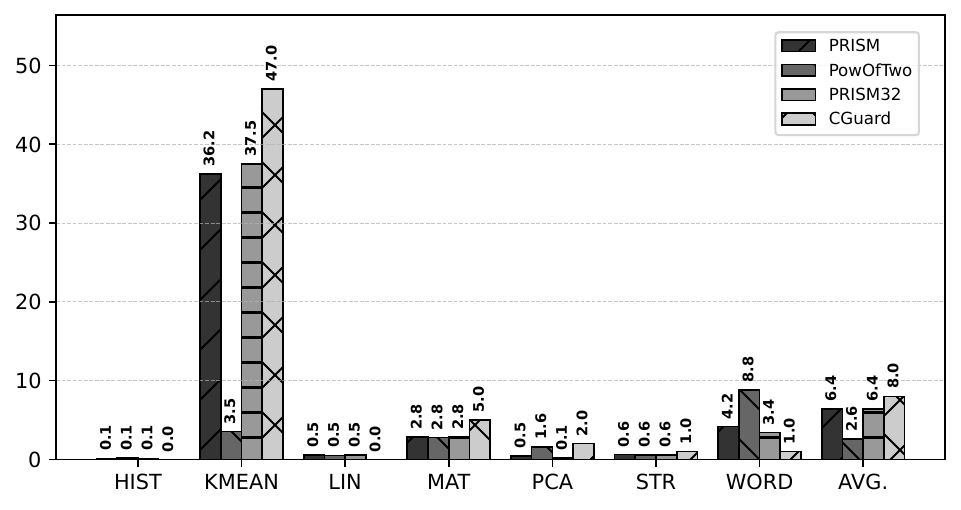}
    \caption{\% increase in peak memory consumption on a single for PRISM, Pow2, PRISM32, and CGuard.}
    \label{fig:phoenix_mem1}
  \end{subfigure}
  \hfill
  \begin{subfigure}[t]{0.48\textwidth}
    \centering
    \includegraphics[width=\linewidth,height=2.5cm]{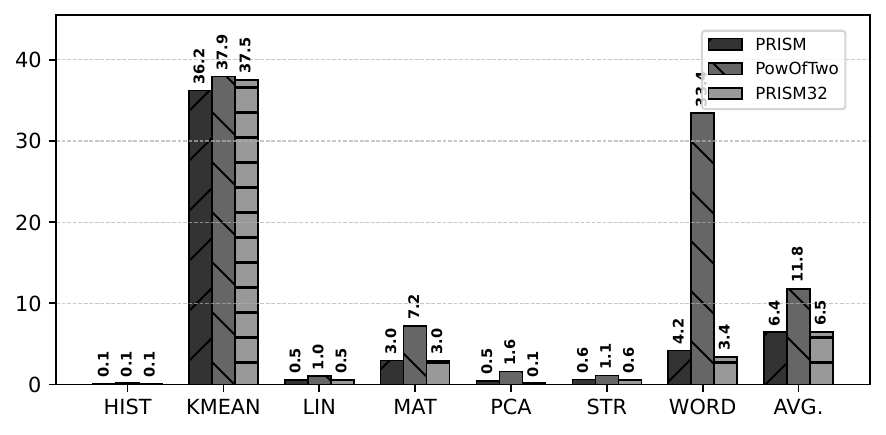}
    \caption{\% increase in peak memory consumption a single processor for PRISM(8), Pow2(8), and PRISM32(8).}
    \label{fig:phoenix_mem2}
  \end{subfigure}


	 \vspace{0.3em} 
  \begin{subfigure}[t]{0.48\textwidth}
    \centering
    \includegraphics[width=\linewidth,height=2.5cm]{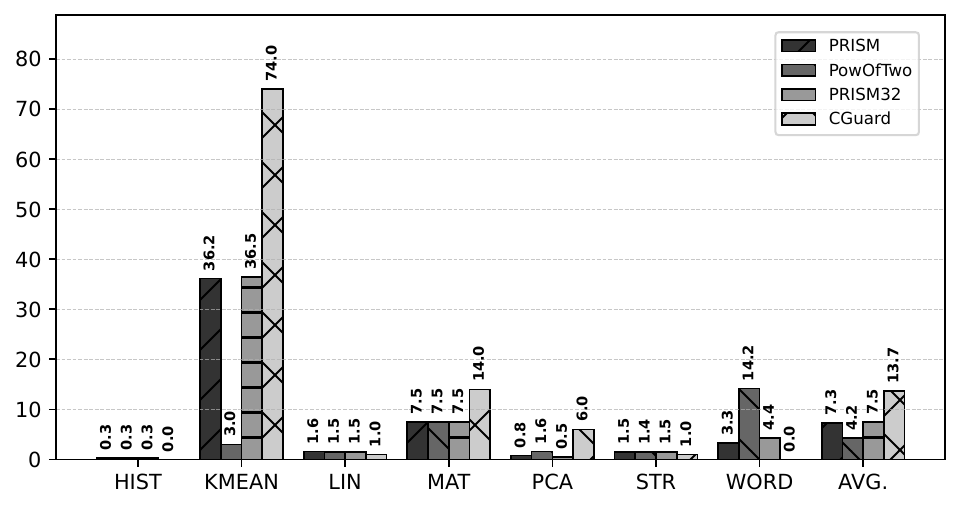}
    \caption{\% increase in peak memory consumption on four processors for PRISM, Pow2, PRISM32, and CGuard.}
    \label{fig:phoenix_mem5}
  \end{subfigure}
  \hfill
  \begin{subfigure}[t]{0.48\textwidth}
    \centering
    \includegraphics[width=\linewidth,height=2.5cm]{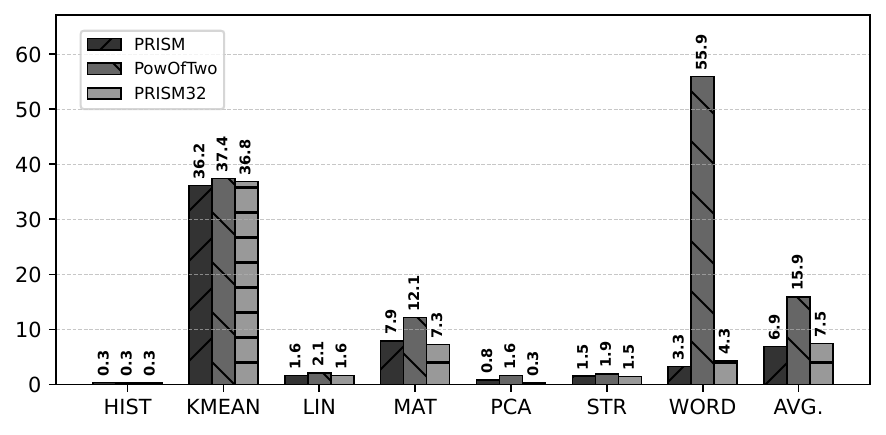}
    \caption{\% increase in peak memory consumption on four processors for PRISM(8), Pow2(8), and PRISM32(8).}
    \label{fig:phoenix_mem6}
  \end{subfigure}
	 \vspace{0.3em} 
  \begin{subfigure}[t]{0.48\textwidth}
    \centering
    \includegraphics[width=\linewidth,height=2.5cm]{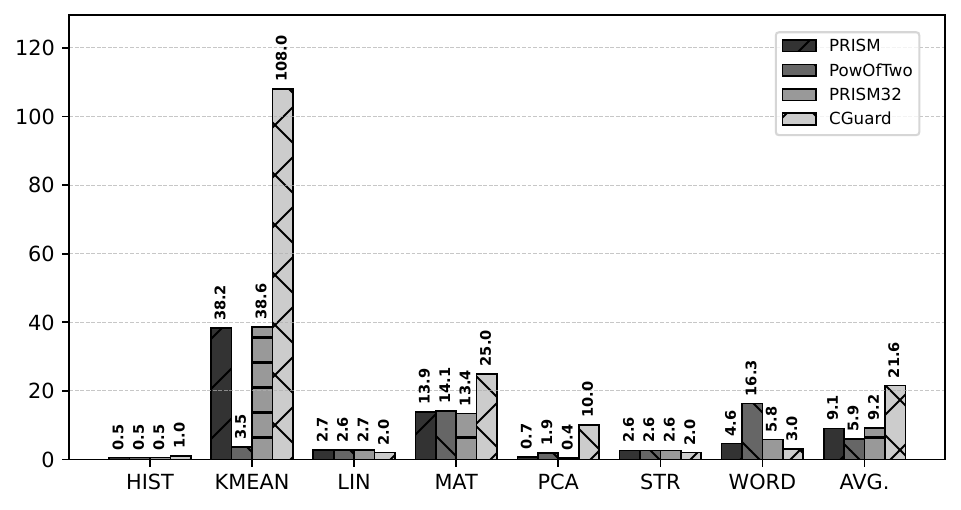}
    \caption{\% increase in peak memory consumption on eight processors for PRISM, Pow2, PRISM32, and CGuard.}
    \label{fig:phoenix_mem7}
  \end{subfigure}
  \hfill
  \begin{subfigure}[t]{0.48\textwidth}
    \centering
    \includegraphics[width=\linewidth,height=2.5cm]{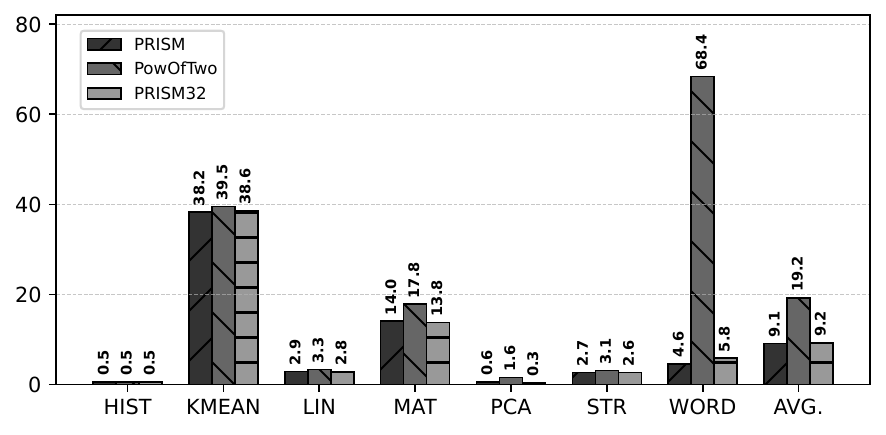}
    \caption{\% increase in peak memory consumption on eight processors for PRISM(8), Pow2(8), and PRISM32(8).}
    \label{fig:phoenix_mem8}
  \end{subfigure}
  \caption{Phoenix memory overhead}
  \label{fig:phoenix_mem}
\end{figure}

For scalability experiments, we ran the Phoenix benchmarks on different numbers of processors. Figures~\ref{fig:phoenix_cpu1}, \ref{fig:phoenix_cpu5}, and~\ref{fig:phoenix_cpu7} show the runtime overhead on one, four, and eight CPUs. Figures~\ref{fig:phoenix_cpu2}, \ref{fig:phoenix_cpu6}, and~\ref{fig:phoenix_cpu8} show the runtime overhead with 8-byte padding on one, two, four, and eight CPUs.
For these benchmarks, we didn't observe a substantial change in the performance with high padding; therefore, we omitted those results.
kmeans, matrix-multiply, and word-count scaled nearly linearly with the number of processors, and therefore our overheads remain similar on all CPUs, except word count. In word-count, Pow(2) exhibits a sharp increase in overhead as the number of CPUs grows. This is because the runtime of this benchmark is just .38 seconds on 8 CPUs in the native run. The memory overhead of this benchmark is increasing with the number of CPUs (see Figure~\ref{fig:phoenix_mem}). Due to the small runtime and high allocation time overhead on large CPUs, the overall overhead increases with the number of CPUs. The reason behind the high memory overhead is due to the 8-byte padding, which alters the allocation buckets, as seen in the case of {\tt nab}. In this case, a 128MB allocation is promoted to the 160MB bucket due to 8-byte padding. Other objects were promoted too, which caused high overheads due to the smaller runtime.

PRISM and Pow2 have similar performance. The overhead of kmeans is between 100-115\% for PRISM, which is lower than the 148\% overhead of CGuard. For histogram, the maximum overhead of PRISM is 30\%, which is lower than the 42\% overhead of CGuard. For PRISM(8), the overheads of kmeans are in the range of 85-106\%, which is lower than PRISM. Pow2(8) also shows similar improvement. The padding doesn't improve the performance of histogram. PRISM32 shows a slight improvement for pca. We found that this was due to the allocator changes. When we ran with just the allocator changes, we got similar results.

Figures~\ref{fig:phoenix_mem1}, \ref{fig:phoenix_mem5}, and~\ref{fig:phoenix_mem7} show the memory overhead on one, four, and eight CPUs. Figures~\ref{fig:phoenix_mem2}, \ref{fig:phoenix_mem6}, and~\ref{fig:phoenix_mem8} show the memory overhead using the 8-byte padding on one, four, and eight CPUs. The peak memory set of kmeans is just 9 MB. In kmeans 99.8\% of the objects were of size 12. In Pow2 mode, we allocate one additional byte and align it to a power of 2. The resulting size is 16. Therefore, in both native and Pow2 mode, these objects were allocated from a bucket size of 16. For PRISM, we added eight additional bytes to store the starting address. These 24-byte objects are allocated from a bucket size of 32. Thus, the memory overheads of both PRISM and PRISM32 are around 37\%. We got a similar overhead if we allocate eight additional bytes during the native run. The overheads of Pow2(8), PRISM(8), and PRISM are nearly the same for kmeans, because the majority of objects were allocated from bucket 32. CGuard reported 47\% overhead for kmeans.

For matrix-multiply, we observed an increase in memory footprint with respect to the number of processors. We found that this is because of our calloc implementation. Because we need to allocate eight additional bytes, we used malloc to allocate the memory and reset it before returning to the user. The default calloc implementation could eliminate the need for resetting memory by using the {\tt mmap} property. In the native run, if we always reset the memory in calloc, we get a similar memory overhead. CGuard also observed the same behavior for matrix multiply. The overheads of Pow2(8) for word-count are very high due to a change in the allocation buckets of some large objects.

\begin{figure}[t]
  \centering
  \begin{subfigure}[t]{0.5\textwidth}
    \centering
    \includegraphics[width=\linewidth,height=2.5cm]{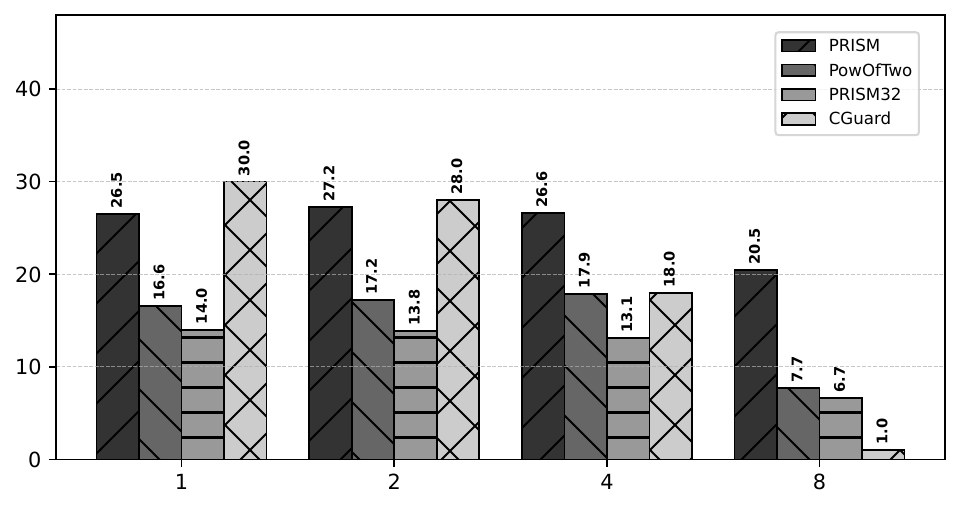}
    \caption{\% decrease in requests per second for PRISM, Pow2, PRISM32, and CGuard on 1, 2, 4, and 8 CPUs}
  \end{subfigure}
  \hfill
  \begin{subfigure}[t]{0.48\textwidth}
    \centering
    \includegraphics[width=\linewidth,height=2.5cm]{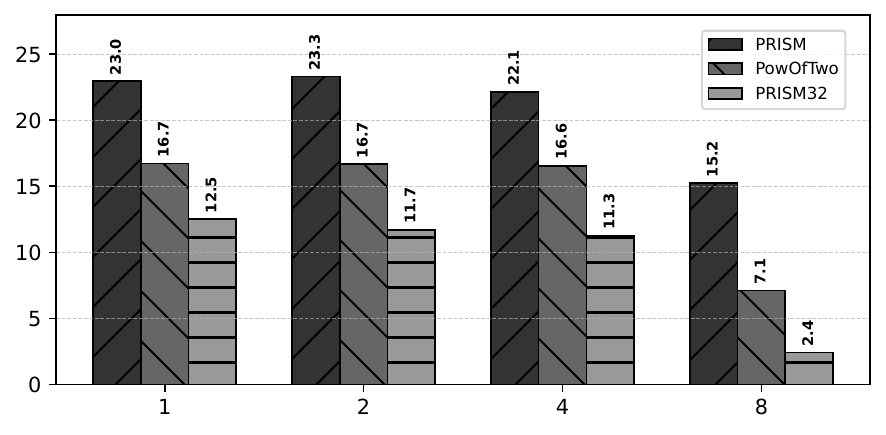}
    \caption{\% decrease in requests per second for PRISM(8), Pow2(8), PRISM32(8) on 1, 2, 4, and 8 CPUs}
  \end{subfigure}
	 \vspace{0.3em} 
  \begin{subfigure}[t]{0.48\textwidth}
    \centering
    \includegraphics[width=\linewidth,height=2.5cm]{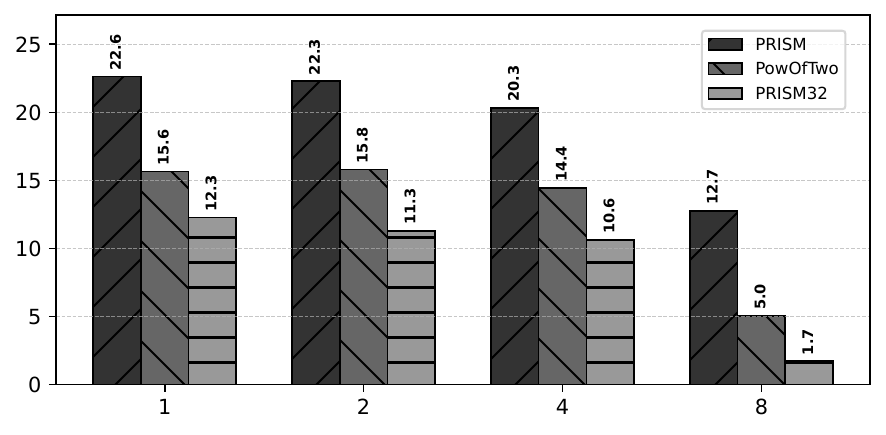}
    \caption{\% decrease in requests per second for PRISM(16), Pow2(16), PRISM32(16) on 1, 2, 4, and 8 CPUs}
  \end{subfigure}
  \hfill
  \begin{subfigure}[t]{0.48\textwidth}
    \centering
    \includegraphics[width=\linewidth,height=2.5cm]{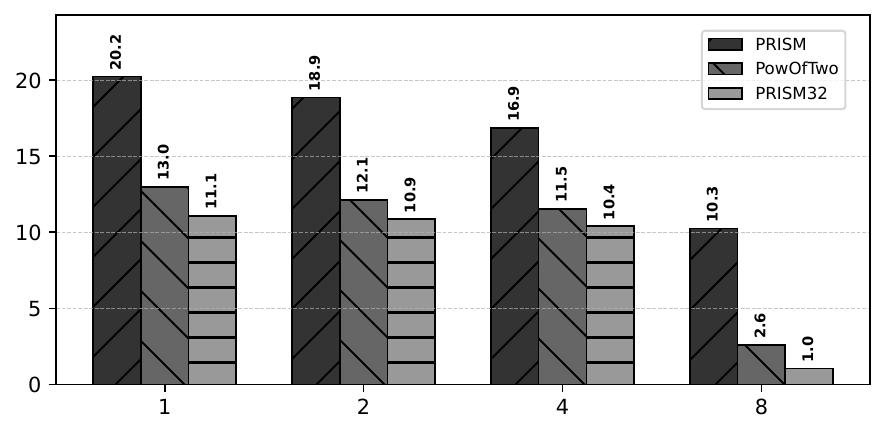}
    \caption{\% decrease in requests per second for PRISM(32), Pow2(32), PRISM32(32) on 1, 2, 4, and 8 CPUs}
  \end{subfigure}
  \caption{Apache CPU overhead}
  \label{fig:apache}
\end{figure}

We evaluated the performance of Apache with different numbers of CPUs, similar to \cite{cguard, baggy}. On eight CPUs, our network card could not fully saturate all eight CPUs, although configurations with fewer CPUs were fully saturated. For each configuration, we varied the concurrency levels until the observed packet drops or a reduction in throughput. We use default pages and use the keep-alive option in the requests. We use requests per second to report our overheads. The requests per second are 22670, 47740, 83450, and 147737 for 1, 2, 4, and 8 CPUs during the native execution, demonstrating near-linear scaling. Figure~\ref{fig:apache} shows the overhead for one, two, four, and eight CPUs. The maximum decrease in throughput for PRISM is 27.2\% on two CPUs. The throughput improved with padding. On a single CPU, PRISM(32) overhead is 20.2\%, resulting in a 6.3\% improvement in throughput compared to PRISM on a single CPU. Pow2 and PRISM32 performed much better than PRISM. The overheads of Pow2 and PRISM32 are 17.9\% and 14\%, respectively, which improve to 13\% and 11.1\% with 32-byte padding. This demonstrates that the q-padding optimization provides a low-overhead solution for enforcing spatial safety in real-world benchmarks. CGuard's maximum overhead is 30\%, which is higher than that of PRISM. Its overhead decreases on four CPUs because, in their setup, the system remained fully saturated only up to three cores.

\section{Security Evaluation and Usability}
\label{sec:security}
\begin{table}
    \centering
    \begin{tabular}{|p{0.20\textwidth}|p{0.24\textwidth}|p{0.5\textwidth}|}
        \hline
      Suite &   Benchmark & Bug detected at \\
         \hline     
   &		bc  &  bc.c:1434, util.c:577, storage.c:177,199 \\
		                      &  gzip & gzip.c:828 \\
	BugBench												&  man & man.c:977,983,155 manfile.c:243 \\
													& polymorph & polymorph.c:120,44,193,194,198,231 \\
													& ncompress & compress42.c:896 \\
         \hline     
		CVE-2013-2028 			  & nginx-1.4.0     & ngx\_recv.c:136, ngx\_http\_request\_body.c:668 \\
         \hline     
		CVE-2011-4971					& memcached-1.4.4     & memcached.c:3534 \\
         \hline     
				SPEC CPU 2017                  & gcc             & tree-ssa-sccvn.c:3365, reload1.c:1868 \\
         \hline     
				SPEC CPU 2017                 & x264             & context\_ini.c:90,91 \\
         \hline     
				Phoenix                       & string           & string\_match.c:158 \\
         \hline     
		    \end{tabular}
    \caption{Bugs discovered in various benchmarks, along with their corresponding source code locations.}
    \label{tab:bug}
\end{table}

To test the effectiveness of PRISM in detecting real-world vulnerabilities, we reproduced the bugs from BugBench as well as the known vulnerabilities CVE-2013-2028 (nginx) and CVE-2011-4971 (memcached). Table~\ref{tab:bug} shows the applications and the program points where PRISM reports the bugs.

PRISM detects all bugs in BugBench. In {\tt polymorph.c} of BugBench, if we continue execution after detecting the first out-of-bounds write, it corrupts a global offset table slot and causes the program to crash. To reproduce subsequent bugs, we insert a dummy global variable after the target to prevent the OOB write from corrupting other program memory.

In nginx, a negative value is passed to the recv system call, which treats it as a large unsigned integer. This causes an attacker-controlled overflow of the server's buffer. Interestingly, if we allow execution to continue after the recv call, PRISM reports the bug again at ngx\_http\_request\_body.c:668, where the program attempts to store an out-of-bounds end address of the buffer.
In memcached, the length argument to memmove is negative, which is interpreted as a large unsigned value.

PRISM also detects all the bugs reported by \cite{cguard} in the SPEC and Phoenix benchmarks. Additionally, it uncovers one more bug in gcc, as described below.

\begin{lstlisting}
// original
rhs2 = nary->op[1];
// corresponding fix
rhs2 = (nary->length > 1) ? nary->op[1] : NULL;
\end{lstlisting}

In this bug, \texttt{nary} has type \texttt{struct vn\_nary\_op\_s} and the eight-byte field \texttt{op[1]} resides at a fixed offset of 40 bytes. {\tt gcc} partially allocates this structure depending on the value of the length field. To eliminate this bug, we applied the fix as shown at line 4. Note that because this memory access occurs at a constant offset from the KSA (\texttt{nary}), PRISM can't detect the violation when q-padding $\ge$ 48. In contrast, to support the size-invariant optimization efficiently, \cite{cguard} manually pads object sizes to be a multiple of their type size. As a result, this bug remained undetected by \cite{cguard}.

Another bug in reload1.c of \texttt{gcc} (shown below) dereferences the memory using an index \texttt{r}. In this case, the dereference offset is not constant with respect to the KSA (\texttt{hard\_regno\_nregs}); therefore, PRISM always detects this bug, regardless of the q-padding value.
In this bug, \texttt{r} may be negative; the fix is shown below.

\begin{lstlisting}
// original
int nregs = hard_regno_nregs[r][PSEUDO_REGNO_MODE (reg)];
// corresponding fix
int nregs = (r < 0)? 0 : hard_regno_nregs[r][PSEUDO_REGNO_MODE (reg)];
\end{lstlisting}

In x264, out-of-bounds global addresses are passed to a routine.
In string\_match.c, the program accesses one byte starting at the end address.

\begin{table}
    \centering
    \begin{tabular}{|p{0.12\textwidth}|p{0.4\textwidth}|p{0.4\textwidth}|}
        \hline
      			Benchmark & Source code modifications & Type of modification \\
        \hline
						perlbench &   util.c:157,278,392,446    & out-of-bounds address computation \\
						gcc       &   ggc-page.c:571,603,628    & performace          \\
						gcc       &   obstack.h:295             & out-of-bounds address computation \\
						x264      &   analyse.c:1099, slicetype.c:258,259 & pointer arithmetic on NULL \\
         \hline     
		    \end{tabular}
    \caption{Source code modification required by PRISM.}
    \label{tab:mod}
\end{table}

Apart from the fixes required for actual out-of-bounds access violations, PRISM require source-level changes when an address that falls outside the inclusive range of [SA, EA] escapes the static scope. These changes were minor compared to the extensive code changes needed by \cite{cguard} to support size-invariant optimization.

Table~\ref{tab:mod} shows the benchmarks, the modified program points, and the type of fix applied. Interestingly, in gcc, performance depends on the top 17 bits of the pointer, which necessitates source-level changes.

In x264, all three modifications involve computing the address of an element pointed to by a structure field; however, the value of the field at runtime was \texttt{NULL}, which resulted in out-of-bounds pointer arithmetic.

The default implementation of \texttt{obstack\_free} in gcc stores the difference of two objects and later recreates the first object by adding the difference to the address of the second object, and passes it to \texttt{obstack\_free}. This causes an out-of-bounds address computation error. To resolve this issue, we switched to an alternative implementation of the obstack API in the same file, which avoids this problem.

In {\tt gcc-page.c}, gcc implements a page table lookup. In 64-bit mode, it creates a linked list node for every unique top 32-bit value in the virtual addresses. Because PRISM metadata is stored in the top 17 bits, the number of linked-list nodes is significantly higher, resulting in high runtime overhead. Note that in PRISM, two addresses on the same page may have different top 32 bits (e.g., say 0x4000400000000 and 0xC000400000010) and thus two nodes are created instead of one, as in the case of native run. To fix this, we use a different mask to reset the top 17-bit from the address during the computation of top 32-bit. This fix required changing only three lines.
For PRISM32, we set \texttt{HOST\_BITS\_PER\_PTR} to 32, which effectively ignores the top bits. We observed a slight performance gain in runtime for PRISM32 due to this change.

In perlbench, the generated addresses are in the inclusive range of [SA-1, EA+32]. However, the memory accesses are always within the bounds. We fixed this by allocating an additional 40 bytes and adding eight and 32 bytes of padding before and after the object, respectively.

\section{Related work}

Spatial memory safety has been widely studied, leading to numerous mechanisms.

Jones and Kelly \cite{jones} first proposed object bounds protection using a splay tree. CRED \cite{cred} extended this to support out-of-bounds pointers.

BaggyBounds \cite{baggy}, PAriCheck \cite{pari}, and LowFatPointers \cite{lfp1, lfp2} constrain object sizes to $2^k$ alignment and use $k$ for fast bounds checks via `XOR'. BaggyBounds and PAriCheck allow pointers beyond EA. We adopted the LowFatPointers design, adding one byte of padding to support EA.

SGXBounds \cite{sgxbounds} limits the address space to 32 bits and stores EA in tag bits. SA is stored at location EA. PRISM follows this model but forbids addresses beyond EA to optimize lower-bound checks. FRAMER \cite{framer} supports a 48-bit space using $2^{15}$-aligned frames for small objects but suffers high overhead due to slow handling of large objects.

CGuard and ShadowBounds provide precise checking but impose type-size constraints. ShadowBounds aborts on violations and lacks protection for stack and global variables, while CGuard traps invalid accesses via tag bits, requiring significant source modifications and suffering from slow trap handling. PRISM addresses these limitations.

CAMP \cite{camp} uses tcmalloc's span-based allocator to infer bounds from size classes. It achieves good performance by enforcing the size-invariant property \cite{cguard} on every pointer arithmetic operation. However, it does not support the EA or partial struct accesses. Furthermore, the granularity of its bounds checks is limited to the size-class of the tcmalloc allocator, rather than the object's actual, precise bounds.

DeltaPointers \cite{delta} uses tagged pointers to detect inter-object overflows. The maximum size of the object depends on the available tag bits.

AddressSanitizer \cite{address} and RangeSanitizer \cite{range} detect overflows and underflows via red zones but not inter-object jumps. AddressSanitizer uses shadow memory; RangeSanitizer relies on aligned allocations. TAILCHECK \cite{tail} improves AddressSanitizer via page protection; however, it can only protect objects of size up to 64 KB and misses underflows.

SoftBound \cite{softbound} tracks per-pointer metadata through all pointer operations and assignments. It provides fine-grained protection, including sub-object overflows, but incurs high memory and CPU overhead.

\section{Conclusion}
We proposed PRISM, a fast and precise object-bounds protection for C programs. PRISM aligns with the defined behaviour of pointer arithmetic in C, ensuring soundness without violating language semantics. The changes required for the correct execution of the SPEC benchmarks are minimal and necessary only to address undefined pointer arithmetic operations. We further showed that allowing programs to access a small, constant offset from a statically known starting address improves the performance of both PRISM and alignment-based schemes, enabling PRISM to match the performance of alignment-based approaches. Further restricting the address space to 32-bit, PRISM achieves a 16.4\% reduction in runtime overhead compared to alignment-based approaches.

\bibliographystyle{plain}
\bibliography{main}

\end{document}